\documentclass[12pt,letterpaper,oneside]{article}
\usepackage[numbers]{natbib}
\usepackage{graphicx}
\usepackage{epsfig}

\def\mbh{$M_{\rm BH}$\/}
\def\nh{$n_{\mathrm{H}}$\/}
\def\lledd{$L/L_{\rm Edd}$}

\def\nc{$N_{\rm c}$\/}
\def\rfe{$R_{\rm FeII}$}

\def\feiiq{\rm Fe{\sc ii}$\lambda$4570\/}
\def\msol{M$_\odot$\/}

\def\rg{R$_{\rm g}$\/}
\def\ltsima{$\; \buildrel < \over \sim \;$}
\def\ltsim{\lower.5ex\hbox{\ltsima}}  
\def\gtsima{$\; \buildrel > \over \sim \;$}

\def\gtsim{\lower.5ex\hbox{\gtsima}}

\def\lya{{ Ly}$\alpha$}
\def\civ{{\sc{Civ}}$\lambda$1549\/}

\def\cmq{cm$^{-2}$\/}
\def\cm3{cm$^{-3}$\/}
\def\hb{{\sc{H}}$\beta$\/}

\def\hbbc{{\sc{H}}$\beta_{\rm BC}$\/}

\def\hbnc{{\sc{H}}$\beta_{\rm NC}$\/}
\def\mgii{{Mg\sc{ii}}$\lambda$2800\/}

\def\ciii{{\sc{Ciii]}}$\lambda$1909\/}
\def\oiiiopt{{\sc{[Oiii]}}\-$\lambda\lambda$\-4959,\-5007\/}
\def\o4363{{\sc{[Oiii]}}$\lambda$4363\/}

\def\heiiuv{He{\sc{ii}}$\lambda$1640}
\def\nv{{N\sc{v}}$\lambda$1240}

\def\feii{{Fe\sc{ii}}\/}

\def\fe{{\sc{Fe}}\/}
\def\gs{{$\Gamma_{\mathrm{soft}}$\/}}
\def\dvr{{$\Delta$v$_\mathrm{r}$}}
\def\vr{{$v_{\mathrm r}$}}

\def\fe76087{{\sc [Fe vii]}$\lambda$6087\/}
\def\oiii{{\sc [Oiii]}$\lambda$5007}

\def\kms{km~s$^{-1}$}

\def\ergss{ergs s$^{-1}$\/}

\def\apj{ApJ}
\def\apjl{ApJL}
\def\apjs{ApJS}
\def\aj{AJ}
\def\apss{ApSpSci}
\def\mnras{MNRAS}\def\pasp{PASP}
\def\aap{AAp}

\title{Quasar  Outflows\\ {\small in the 4D Eigenvector 1 Context}\footnote{To appear in the {\em Astronomical Review}.}}
\author{Paola Marziani\footnote{INAF, Osservatorio Astronomico di Padova, Italia and Instituto de Astrofisic\'{\i}a de Andalus\'{\i}a (CSIC), Spain.}, and  Jack W. Sulentic\footnote{Instituto de  Astrof\'{\i}sica  de Andalus\'{\i}a (CSIC), Spain.} }
\begin{document}
\maketitle





\begin{abstract}
Gas outflows appear to be a phenomenon shared by the vast majority of quasars. Observations indicate that there is wide range in outflow prominence.  In this paper we review how the 4D eigenvector 1 scheme helps to organize  observed properties  and lead to  meaningful constraints on the outflow physical and dynamical processes.
\end{abstract}

\section{Introduction: organizing quasar diversity}

It is perhaps not surprising that quasars were thought to be 
spectroscopically similar in the early days of quasar research. 
Accretion phenomena  are rather mass invariant so no great diversity was 
expected and available low s/n spectra certainly looked rather similar.
Accretion luminosity can be written as $L_\mathrm{acc} \propto M \dot{M} / R$, 
where $R$\ indicates distance from the central black hole customarily identified 
with the radius of the last stable orbit, $\sim$ \rg. It follows that 
$L_\mathrm{acc} \propto \dot{M}$: so the energetic output is independent of 
the black hole mass \mbh.  

Recent times have seen more attention given to spectral differences that may be
related to the wide range of Eddington ratio likely to be present in the quasar 
population now much better sampled via optical surveys (bright quasar survey (BQS), Hamburg-ESO (HE), Sloan digital sky survey (SDSS), etc.; 
\cite{sulenticetal00a,baskinlaor05b,yipetal04,kuraszkiewiczetal09}). A landmark study of systematic trends 
in quasar spectra  \cite{borosongreen92} involved a principal component analysis (PCA) 
of 87 bright quasars from the Palomar-Green survey. Correlations among several 
parameters measured for emission features in the \hb\ spectral region  could be 
grouped into an Eigenvector 1 (E1) involving, among other things,  an anti-correlation 
between \oiiiopt\ and optical \feii\  strength. Variations of E1 were later found in a 
number of larger samples 
\cite{marzianietal96,boroson02,grupe04,kuraszkiewiczetal02,sulenticetal00a,sulenticetal02,yipetal04,kovacevicetal10,kruzceketal11,tangetal12,wangetal06}.  The E1 discovery was eventually generalized into a 4D Eigenvector 1 parameters (4DE1) space \citep{sulenticetal07} which added  measures of the soft X-ray photon index (\gs) and  \civ\   broad line profile shift. The higher dimensionality of the input space is helpful to cover parameters that are independent 
(``orthogonal,'' one is tempted to say) observationally and that correspond to different physical processes: 1) the dynamics of the low-ionization line (LIL) emitting gas, 2)  physical conditions and metallicity of the line emitting gas, 3) sources of continuum emission and 4) the dynamics of the high ionization line (HIL) emitting region.

It is now clear that all type-1 AGNs do not show similar (emission line) spectra anymore than 
all stars show similar (absorption line) spectra. Differences go beyond source-to source comparisons
and include significant differences between median spectra of large numbers binned in a context 
like 4DE1. We now suspect that Eddington ratio,  central black hole mass,  orientation and metallicity 
-- parameters that could define a 4D physical space related to accretion properties of the central 
supermassive black hole, the process that is believed to power the tremendous energy output of 
quasars -- all likely affect the observed source spectra.

\begin{figure}[ht]
\centering
\includegraphics[width=5in]{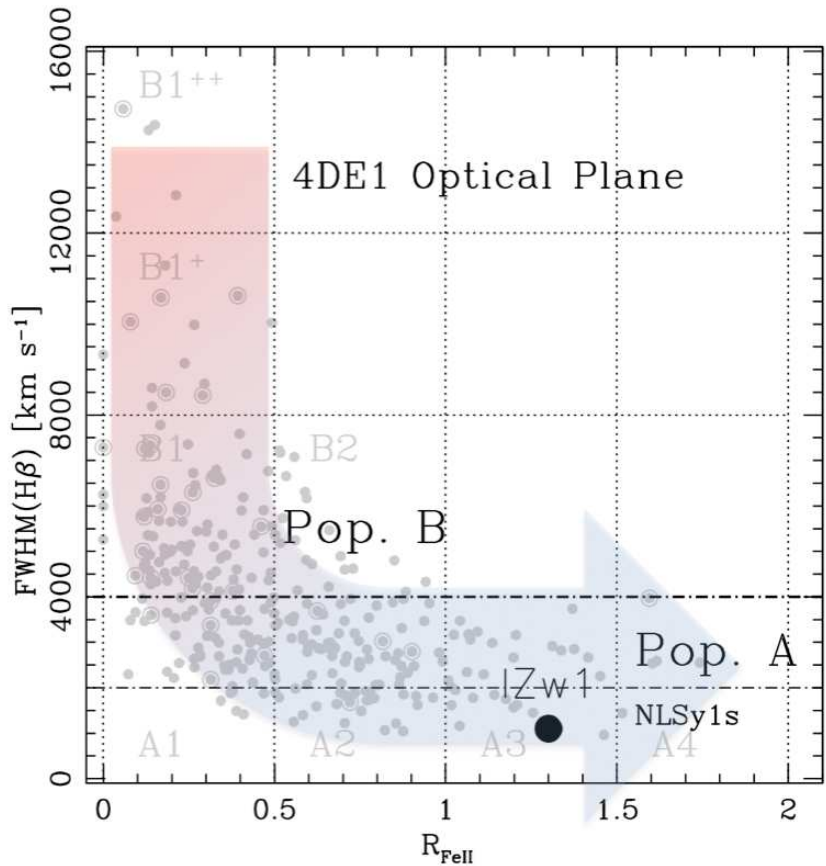}
\caption{The 4DE1 optical plane. In abscissa the ratio \rfe\ as defined in the text, and in ordinate the FWHM of the broad component of \hb. Bins in the plane define spectral types. The arrows schematically indicates the increasing prominence of blue shifts (outflows) along the elbow-shaped sequence. The most compelling evidence of outflows is observed in the extreme Pop. A bins, A3 and A4. }
\label{fig:optplane}
\end{figure}

The 4DE1 formalism has reduced interpretative confusion and assisted  the interpretation of several  
observational aspects that appear puzzling if, for example, sets of spectra are indiscriminately 
averaged together \citep{nagaoetal06}. Spectra can be averaged but only in a well defined context like 4DE1.
An obvious indication that not all quasars are the same in terms of physical and dynamical properties comes
from the difference between optical spectral properties of narrow line Seyfert 1 sources (NLSy1s)
(which can be luminous quasars) and radio-loud quasars (a less extreme difference is shown in Fig. 2 
of \cite{sulenticetal00a}). In the former we observe  narrow low ionization lines dominated 
by \feii\ emission vs. broad multi-component (redward asymmetric) lines in an overall higher 
ionization spectrum. The systematic differences are recognized by several authors who speak, emphasizing different 
aspects, of NLSy1s and broad(er) line quasars (BLQs), of Population A and B \cite{sulenticetal00a}, Population 1 and 2 \cite{collinetal06} or 
disk- and wind-dominated sources \cite{richardsetal11}. In this review we adopt the empirically-motivated Pop. A and 
B subdivision involving FWHM(\hbbc) $\approx$ 4000 \kms. Pop. A and B are, respectively,
narrower and broader than this value. The reason for this limit is made clear by Fig. 
\ref{fig:optplane}.

The division into two populations is  useful for highlighting major differences among Type 1 AGNs, although spectral differences among objects within the same population are still noticeable, especially for Pop. A sources (see Figure 2 of \cite{sulenticetal02}). This motivates gridding the 4DE1 optical plane into bins of FWHM\-(\hbbc) and iron emission strength.   The prominence of \feii\ emission is quantified  the intensity of the \feiiq\ blend of multiplets normalized to the one of \hbbc: \rfe = \feiiq/\hbbc. Bins A1, A2, A3, A4 are defined in terms of increasing \rfe, while bins B1, B1$^{+}$, and B1$^{++}$\ with bin size $\Delta$\rfe  = 0.5,  are defined in terms of increasing FWHM(\hbbc).  Sources belonging to the same spectral type show similar spectroscopic measures (e.g., line profiles and line flux ratios). They are assumed to isolate sources with similar broad line physics and geometry.  Systematic changes are minimized (although they may not be fully random \cite{marzianietal01}) within each spectral type so that an individual quasar can be taken as representative of all sources within a given spectral bin. The binning adopted in \cite{sulenticetal02} (see Fig. \ref{fig:optplane}) has been derived for low-$z$ ($<$0.7) quasars and is luminosity dependent since the FWHM boundary between Pop. A and B is  luminosity dependent  \cite{marzianietal09,dultzinetal11}.

In this paper we concentrate  on a particular physical process, outflows in or in the proximity of the broad line region (BLR) that show a clear trend along the 4DE1 sequence. We will focus on observational constraints to the interpretations about this process. Quasar outflows may have important effects on the host galaxy and perhaps also on the intergalactic medium.  A comprehensive  account can 
be found in Chapters 8 and 9 of the recent book celebrating the fiftieth anniversary of quasar discovery \cite{donofrioetal12}. 

\section{Quasars non-relativistic outflows: an optical/UV perspective}

\paragraph{Setting a proper rest frame } An accurate rest frame definition is needed to evaluate the meaning of inter-line shifts but it remains a sore problem for quasar studies. Knowing the rest frame  makes  possible a physical interpretation of    internal shifts among broad lines that are distributed over a wide range of radial velocity, differ widely  from source to source, and  depend on ionization state \citep{gaskell82,tytlerfan92}. The problem remains especially serious  for high redshift quasars since  prominent narrow emission lines are shifted into the IR, a range that has been of difficult coverage until very recent times. Narrow LIL serve as rest frame reference, and they are assumed to provide a value closest to the one of the host galaxy, although this is not really {\em known} for the wide majority of sources).

Setting a proper rest frame is of special importance since outflows in quasars are suggested by the detection of blueshifted absorption or emission components of spectral lines if shifts are interpreted as due to Doppler effect (for a dissenting view see \cite{gaskell09}). These features can cover a wide range of shifts and widths, and are present in a sizable fraction of quasars. Narrow absorption lines (FWHM $\ltsim$ 1000 \kms) are very frequent in HIL \civ\ \cite{vestergaard03,wildetal08}, and  associated to a wide range of outflow velocities (up to $12000$ \kms).  X-ray emission lines between a few tens eV and 1 KeV in the so-called warm absorber are typically blueshifted by $\sim -1000$ \kms\ and are probably part of a phenomenon involving most if not all quasars \citep{blustinetal05,chakravortyetal12}.  An interesting study could involve the warm absorber prevalence and properties along  the E1 sequence but this study, as far as we know, has not been carried out as yet. 

\begin{figure}[ht!]
\centering
\includegraphics[width=4in]{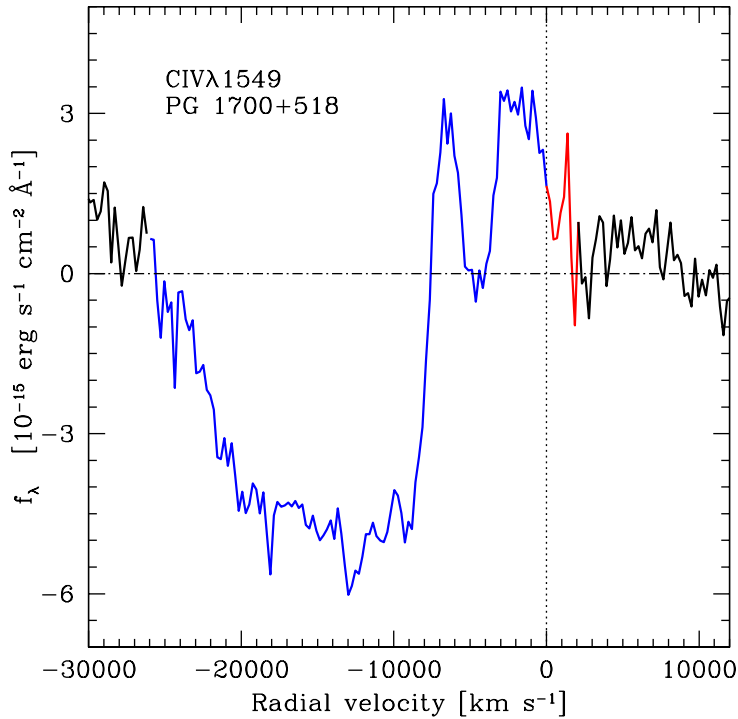}
\caption{The \civ\ profile of the BAL QSO PG 1700+518, extracted from the continuum-subtracted spectrum. Abscissa is redial velocity with respect to rest frame; ordinate is specific flux. The rest frame has been set from \hbnc\ and is therefore believed to be especially accurate. }
\label{fig:bal}
\end{figure}

\paragraph{BAL QSOs} The presence of broad absorption lines (BALs) systematically blueshifted with respect to the emission component readily indicates outflow of gas, and that momentum is being transferred from the quasar continuum to the outflowing gas. The  maximum radial velocity of the absorption throughs can reach several tens of thousands \kms, far above the escape velocity at the distance where they are probably formed. Fig. \ref{fig:bal} shows the \civ\ profile after continuum subtraction of the classical high ionization BAL QSOs PG 1700+518 \cite{sulenticetal06}.   The SDSS survey indicates that BAL QSOs are detected with a frequency of 15-20\%\ in large quasars samples \cite{halletal02,reichardetal03,gibsonetal09}. BAL QSOs were discovered soon after the first quasars \cite{lynds67}, but the fundamental issue of whether they represent a special subsample of quasars or are instead quasars observed at a special viewing angle is still lingering today \cite{gangulybrotherton08}. 

The so-called line-locking (i.e., the presence of absorption trough in \civ\ with the same radial velocity separation of \lya\ and \nv\ indicates   absorption by C$^{+3}$ ions of  down-shifted \lya\ and \nv\  line photons \cite{weymannetal91}. This discovery has oriented eventual modeling of the BAL QSOs toward the inclusion of significant resonant line acceleration \cite{progaetal00}. In this case the outflow is accelerated by the emission of line photons in gas  that continues to see unabsorbed  continuum photons because of its changing velocity with distance from the central continuum source \cite{castoretal75}.

\paragraph{Broad emission lines} The first evidence of a systematic blueshift of the \civ\ line with respect to \mgii\ came in the early 1980s \cite{gaskell82,espeyetal89}. Since then evidence has considerably strengthened with several papers appearing in the 1990s confirming the previous result (the main development until 1999 are summarized by \citet{sulenticetal00a}).  In the  1990s the UV data provided by HST spectrographic cameras made possible a  comparison between low- and high-ionization lines for many tens of low-$z$\ quasars \cite{corbin90,willsetal93,brothertonetal94a}. Quasars HILs were found to be systematically blueshifted with respect to LILs, but not in all cases \citep{marzianietal96}.  

\begin{figure}[ht!]
\centering
\includegraphics[width=4in]{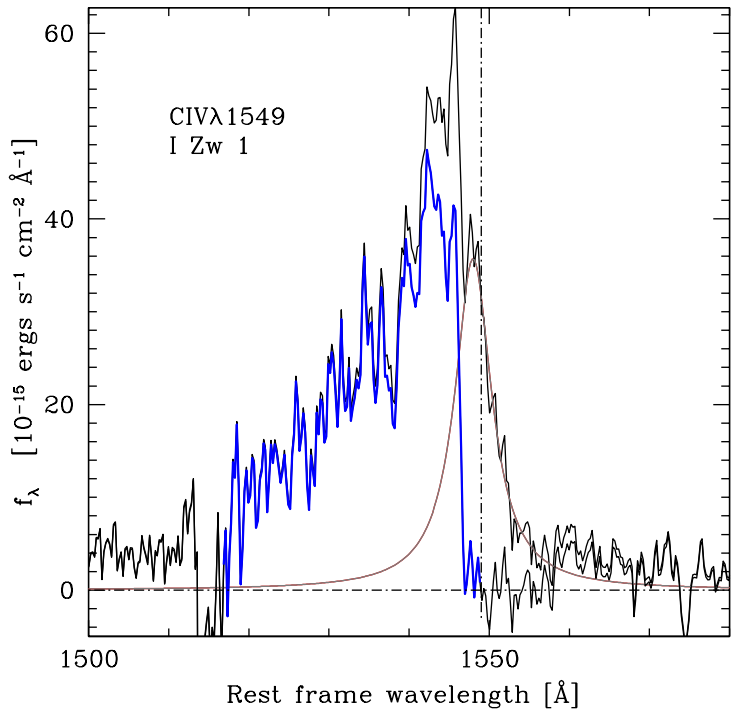}
\caption{The \civ\ profile of the extreme Pop. A source I Zw 1. Abscissa is wavelength in \AA, and ordinate is specific flux. A scaled \hbbc\ profile (thin grey line) has been overlaid to emphasize the difference between the HIL \civ\ and LIL \hb.  }
\label{fig:izw1}
\end{figure}

The systematic blueshift of \civ\ in the low-$z$ NLSy1 source I Zw 1 is an exemplary, albeit extreme case. Superimposing a scaled and shifted broad \hb\ profile to \civ\ shows that the \civ\ can be accounted for by (1) a  component reproduced by the scaled profile of \hb\ plus a fully blueshifted, much broader component that accounts for most of the \civ\ flux. In Fig. \ref{fig:izw1}, a scaled  \hb\ matched component  is superimposed to the original profile of \civ. The interpretation is straightforward in this case: \hb\ is mainly emitted by a flattened distribution of gas seen face-on, or almost so, while the  gas emitting the blueshifted component  is moving toward us. The receding part of the flow is assumed to be hidden by the structure emitting \hb, yielding a net blueshift in the \civ\ profile. I Zw 1 is in many ways not a peculiar source but an extreme in the 4DE1 sequence. Wind emission appears to be dominant for the HIL \civ, and indeed I Zw 1 could be considered a prototype of ``wind-dominated'' quasars \cite{richards12}. Sources like I Zw 1 are clearly unsuitable for estimating the central black hole mass under the assumption of virial motions, unless a correction is applied \cite{sulenticetal07,netzeretal07,marzianisulentic12}.

\section{Outflows along the 4DE1 sequence} 

\begin{figure}[ht!]
\centering
\includegraphics[width=5in]{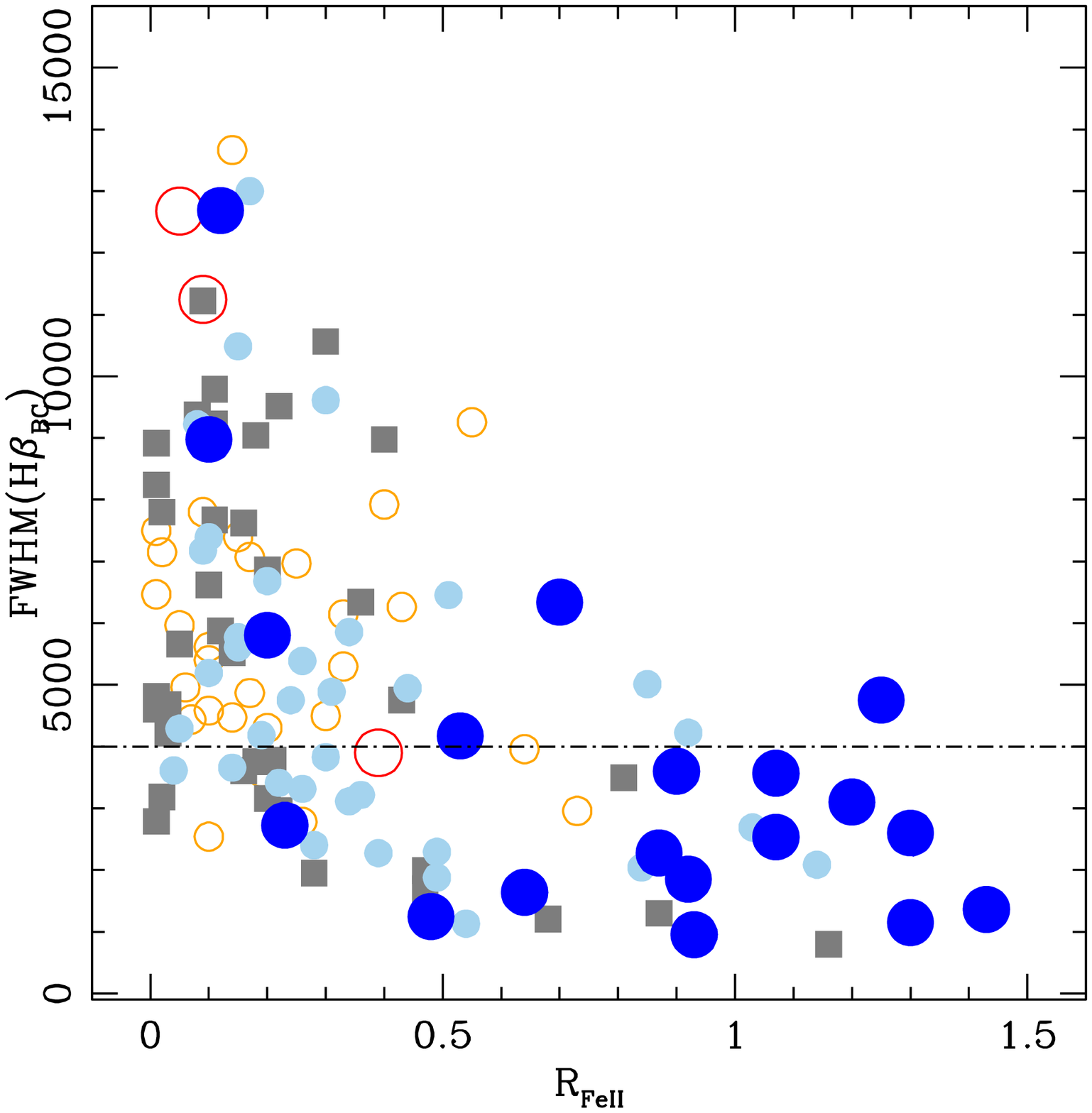}
\caption{\civ\ shifts in the optical 4DE1 plane. Data points from \cite{sulenticetal07} are represented with different symbols according to the centroid at half-maximum shift of the broad \civ\ profile. Large blue: \dvr $<-1000$\kms; pale blue: $-200 $ \kms$>$ \dvr $\ge -1000$\kms; grey (unshifted):   $-200$ \kms $\le$ \dvr $\le 200$\kms; orange open circles: $200< $ \kms \dvr $\le 1000$\kms; red open circles: \dvr $>1000$\kms.}
\label{fig:civshiftse1}
\end{figure}

As one moves along the E1 sequence toward Pop. B, the appearance of \civ\ changes \citep{bachevetal04,sulenticetal07} with large \civ\ blueshifts becoming less frequent (see Fig. \ref{fig:civshiftse1}).   The maximum shift amplitude decreases as well. B1 sources account for the wide majority of Pop. B: their \civ\ and \hb\ profiles are similar, and profile parameters like centroids and asymmetry occupy overlapping ranges \citep{marzianietal96}.  This trend has been recently explained with the progressive demise of the wind component \cite{marzianietal10,richardsetal11,wangetal11}. An interpretation of the BLR  that seems to be supported, at least in part, by micro-lensing of the BLR \cite{sluseetal11,sluseetal12} involves three main regions: (1) a relatively stable region whose dynamics is predominantly virial, associated to a symmetric unshifted emission in emission lines; (2) an outflow region revealed by emission line blueward asymmetries or by a component systematically blueshifted with respect to  the rest frame of the quasars. It has proved useful to introduce a third region in Pop. B: (3) a very-broad line region (VBLR)  that accounts for the innermost emission and for the prominent redward asymmetries observed in the line profile of these sources. The introduction of this third region is motivated by the ``inflected'' appearance of  \hbbc\ line profiles, as well as by the absence of low-ionization line emission. Optical \feii\ emission is not emitted in the broadest component associated to the VBLR. This said, one should be aware that the VBLR cannot be rigorously isolated in the profile, and that there could be, if virial motions account for the bulk line broadening, a smooth gradient of ionization with distance from the central continuum source. Empirically, the most prominent lines can be  modeled by the addition of three components (the VBC is 0 in Pop. A sources), each of them associated to the regions described above. 

A blueshifted component that appears dominant in the \civ\ profile of A3/A4 spectral types is also apparent in \hb, although the intensity ratio is \civ\hb$\gg$1. The \hb\ profile is mostly unaffected \cite{marzianietal10}  when the \civ\ blueshifted component is less prominent. 

The blueshifted component in \civ\ and other high ionization lines seems to be present in most spectral types. Only in spectral type B1$^{+}$\ it is irrelevant and undetectable or almost so in bin B1$^{++}$. These bins are the ones with the largest fraction of RL FRII galaxies \citep{zamfiretal08}.  It appears reasonable that the presence of axial, powerful relativistic outflows may hinder the propagation of a strong disk wind as observed in radio-quiet sources \cite{normanmiley84}. However, recent work suggests that some wind emission is coexisting with relativistic jets \cite{richardsetal11}. The parameter that is more relevant for the prominence of winds appears to be \lledd. Radio-loudness might  yield only a secondary effect since occurrence of powerful radio-loud sources is favored in sources of low \lledd\ \cite{woourry02,marzianietal03b}. 

\subsection{More on outflows in extreme Pop. A sources}

The  distribution of \oiii\  shifts with respect to rest frame (set by low ionization narrow emission lines) is strongly skewed toward the blue \cite{zamanovetal02,huetal08}, indicating that outflow, at some level, is also occurring within the NLR in a sizable fraction of sources. Also, most sources show a typical profile with a spiky, symmetric narrow core and a blueward asymmetry: the latter feature can be easily associated to fast outflowing gas. 

Observations of the extreme Pop. A  I Zw 1 shows an \oiiiopt\ blueshift of $\approx$ --500 \kms\ relative to rest-frame measures\cite{borosonoke87,marzianietal96}, with a second component at $\approx$ --1500 \kms\ (\cite{veroncettyetal04}, and references therein). I Zw 1 is not the sole source showing a systematic blueshift of \oiii.   It is possible to identify I Zw 1 analogues  with  $\Delta v_{\rm r}$(\oiiiopt) $ \gtsim -300$ \kms\ with respect to \hbnc.  These sources are found to be rare
\cite{grupeetal99,grupeetal01,marzianietal03b,komossaetal08,zhangetal11}) and  seem to lie out of a continuous, skewed shift distribution (hence the frequent use of the   ``blue outliers" denomination).  The distribution of blue outliers in the
E1 optical plane  is  different from that of the general quasars
population \cite{zamanovetal02} since they occupy the
lower right part of the diagram and are exclusively Pop. A/NLSy1
nuclei \cite{zamanovmarziani02}. 

It is interesting to point out that the blue outliers have low W(\oiiiopt) at low $z$. Sources with low W(\oiiiopt) appear to show only a blue-shifted \oiiiopt\ profile as if only the asymmetric part of the most typical profile were  present (i.e., the ``spiky'' component were missing). It is relatively straightforward to interpret the large blueshift and the profile of the \oiiiopt\ lines as the result of an outflow \cite{aokietal05,zamanovetal02}, within a compact NLR \cite{veroncettyetal04,zamanovetal02}. Zamanov et al. \cite{zamanovetal02} constructed a  model of gas moving following a wind velocity field (\S \ref{wind}). Both \oiii\ and \civ\ were assumed to arise in a radial flow constrained in a cone of large half-opening angle, where the receding part of the flow is obscured by an optically thick accretion disk. This simple model could explain  the observed profiles of both lines as emitted at different distances within the same flow. The model  consistently suggested a very compact NLR ($r \ltsim$ 1 pc for a black hole mass of 10$^{8}$\ \msol). 

Most extreme examples of blue outliers at high $z$ may involve massive outflows on scales of several kpc  \cite{canodiazetal12}. In this case the receding part of the flow is obscured by the dust and gas in the host galaxy. 

\subsection{Outflows and quasar evolution}

It is  intriguing to note that the observed blueshifts seem to constrain a scale invariant BLR/NLR  structure. In both BLR and NLR we observe  HIL blue shifts, while LIL appear to be more symmetric. Extreme Pop. A  sources that are blue outliers  are outflow dominated  sources within  both the BLR and NLR. It is tempting to speculate that, as the active nucleus evolves, the NLR occupies progressively more space, yielding to the extended NLR that have been spatially resolved in nearby Seyfert galaxies. In this way the E1 sequence appears as an evolutionary sequence: from the fledgling extreme Pop. A sources radiating a large Eddington ratio, wind dominated, to the ``dying'' Pop. B sources that host very massive black hole, radiate at low Eddington ratio and show very extended NLR \cite{sulenticetal00a,mathur00}. \mbh\ and NLR prominence may serve as quantitative evolutionary parameters.  

\subsection{Outflows and the LIL emitting part of the BLR} 

Several works stressed since the mid-1980s that LILs and HILs have to be emitted at least in part from different media \cite{collinsouffrinetal88}.  The effect of the strongest winds may have important effects on the very structure of the BLR \cite{leighly04}. Extreme Pop. A sources apparently have only a very dense low-ionization emitting gas, with little contribution of gas at \nh $\sim 10^{10}$ \cm3 \cite{negrete11,negreteetal12}. Their \ciii\ emission is the lowest observed in quasar spectra. Another speculation would be to consider these sources as experiencing a radiation force strong enough to sweep away/heat the lower density gas within the BLR. It is interesting to note that extreme Pop. A sources still show an unperturbed \hb\ component. A stable equilibrium might however be possible only for the densest gas.

\section{A simple  physical scheme}
\label{wind}

From an observational point of view it is not settled whether we are considering the flow of a continuous medium or the motion of discrete gas clouds. Gas clouds could be magnetically confined \cite{reesetal89}; however, high-resolution spectroscopy fails to resolve individual clouds \cite{aravetal98,laoretal06} placing a lower limit on their numbers. This result is consistent with the idea that we are observing a hydrodynamical flow. Such flows are appealing especially at low luminosity and in the presence of a hard spectrum but are very difficult to test observationally \cite{everett07} even if they have been successful in reproducing line profiles \cite{bottorffetal97,bottorffferland00}. Thermal winds, i.e., winds where the outflow is driven by the thermal energy of the gas, appear unable to withstand cooling losses if they are expanding adiabatically \citep{everettmurray07}.   Radiative acceleration models have been successful in explaining single-peaked profiles assuming the lines are emitted in a disk wind \cite[e.g.,][]{murraychiang97,murraychiang98}. It is, however, not clear to us whether typical Balmer and \civ\ profiles in Pop. B sources can be explained by such a model (although see \citep{flohicetal12}). 

\begin{figure}[ht!]
\centering
\includegraphics[width=5in]{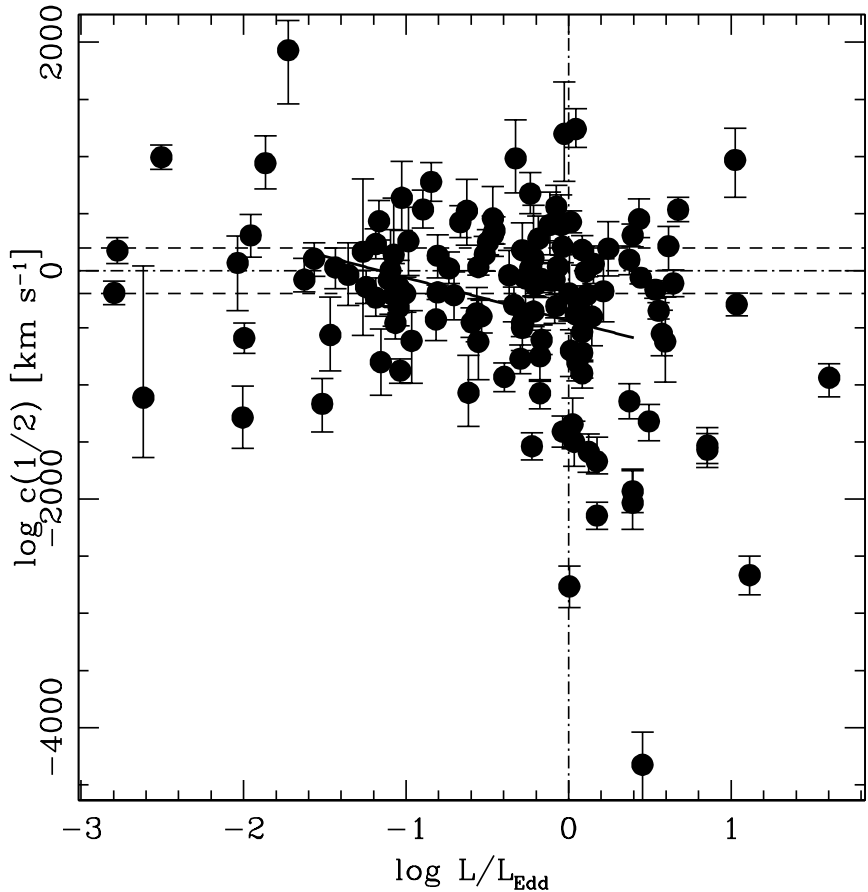}
\caption{\civ\ centroid at half maximum shifts as a function of logarithm of Eddington ratio. Data  from \cite{sulenticetal07}. The dashed lines define a $\pm$ 200 \kms\ radial velocity interval where the \civ\ line profile are considered unshifted. }
\label{fig:civshifts}
\end{figure}

Observationally, the dependence of the outflow blueshifts  on Eddington ratio \cite{gangulyetal07,bachevetal04,baskinlaor05b,sulenticetal07} strongly indicates that radiation forces are at play right in cases where winds/outflows appear to be most prominent. Fig. \ref{fig:civshifts} shows the \civ\ broad component centroid at half maximum (i.e., a measurement of the global \civ\ broad component shift) as a function of \lledd. The simplest interpretation sees a  correlation between negative (i.e. blue) shift amplitude and \lledd\ for sources in the range $-1.6 \le \log$\lledd $\le 0.6$. The large scatter is at least partly due to orientation effects \citep{richardsetal02}. A more proper interpretation might 
involve a  discontinuity at $\log$\lledd $\approx$ -0.5: from a distribution of shifts that is fairly symmetric to a distribution skewed toward negative values, where the largest amplitude shifts are observed. Moderate amplitude blueshifts are not infrequently found also at relatively low \lledd.

The balance of gravitational and radiative forces is likely  different in each quasar (for fixed \mbh) and    dependent on the luminosity-to-mass ratio (i.e., the Eddington ratio).   The ratio between  radiative  and  gravitational acceleration of  an optically thick, Compton-thin cloud of surface $\Sigma$\ can be written as
$
r_\mathrm{a} =  {a_{\mathrm{rad} } }/{a_{\mathrm{grav} } } \approx 
({{\Sigma L_{\mathrm{ion}}}/{4\pi  r^{2} c M_{\mathrm{cloud}}}})/({{GM_\mathrm{BH}}/{r^{2}}})  \approx 7.2 \frac{L}{L_\mathrm{Edd}}  N_\mathrm{c,23}^{-1}$
\citep[cf.][]{netzermarziani10,marzianietal10}.  The momentum transferred to the gas in a cloud of mass  $M_{\mathrm{cloud}}$\ is assumed equal to the total momentum from the ionizing radiation $L_\mathrm{ion}/c$. The relation holds as long as the gas remains optically thick and resonant-line acceleration is negligible.  Opacity effects are ignored. We will therefore  use the relation only to outline a qualitative trend. If there is a distribution of column density within the BLR the lowest density gas can become unbound as the Eddington ratio increases.   Radiative acceleration will dominate if  $r_\mathrm{a} \gg$ 1. This is likely  the case for the blueshifted emission component. Equilibrium may be approached at   $r_\mathrm{a} \approx$ 1. \lledd $\approx$1 (the corresponding \nc\ should be $\ltsim 10^{24}$ \cmq\ which is a plausible value for the LIL-BLR). 
Finally if the ratio  $r_\mathrm{a} \ll$ 1 the emitting gas (in the presence of drag forces or viscous stresses that lead to angular momentum loss) may be unable to withstand the central black hole gravity and may fall toward the center giving rise to the observed redshifted profiles typical of low-Eddington ratio Pop. B sources. The dynamical status of the line emitting gas may ultimately be related to \nc\ and source \lledd.   This scheme readily explains the ``coexistence'' of a gravitationally bound region and a radiatively driven outflow  \cite{denneyetal09b,marzianietal10,wangetal11}. 

What are the physical conditions in the outflowing gas emitting the broad lines? The blueshifted component is prominent  in \lya\ and \civ. It is revealed in a consistent fit of \heiiuv\  blended with \civ, and it appears to be weak (undetected in LILs) implying a rather high  lower limit to \civ/\hb\ and \lya/\hb. The weakness of any LIL emission and the \heiiuv/\civ\ ratio imply  high ionization parameter $U\gtsim$ 10$^{-1}$\ \cite{leighly04,marzianietal10} validating the description ``high ionization-outflow or wind.'' Winds in extreme Pop. A source may disperse a highly enriched gas \cite{wangetal09a}.

\subsection{BAL QSOs} 
\label{bals}


Outflows driven by line and ionizing photon pressure  can accelerate the line-emitting gas to 
$v_\mathrm{t} \approx k \left({\cal M} { L}/{r}\right)^{\frac{1}{2}}  \approx  10^{4} {\cal M}^{\frac{1}{2}} L_{46}^{\frac{1}{2}} r_{16}^{-\frac{1}{2}} \mathrm{km ~~s}^{-1}$ where ${\cal M}$\ is the force multiplier  
 \citep[e.g.][]{laorbrandt02}. 
 The same relation can be written in terms of \lledd\ as $v_\mathrm{t} \approx \left({\cal M} { L}/{L_\mathrm{Edd}}\right)^{\frac{1}{2}} v_\mathrm{Kepl}$\ where $v_\mathrm{Kepl}$\ is the Keplerian velocity at the launching radius of the wind. Therefore it is not necessary for the Eddington ratio to be high in order to launch a wind \cite{laorbrandt02}, provided that ${\cal M} \gg 1$\ as in the case of line driven winds \citep{progaetal00}. BAL QSOs like PG 1700+512 are extreme Pop. A sources radiating at high \lledd.  PG 1700+512 shows a very large $v_\mathrm{t}$\ inferred from the maximum \vr\ in the \civ\ absorption trough.  This extreme phenomenology is probably restricted to highly accreting sources.  However BALs are observed in sources of relatively low Eddington ratio suggesting that line driving is important \cite{sulenticetal06,gangulyetal07}.


\begin{figure}[ht!]
\centering
\includegraphics[width=4in]{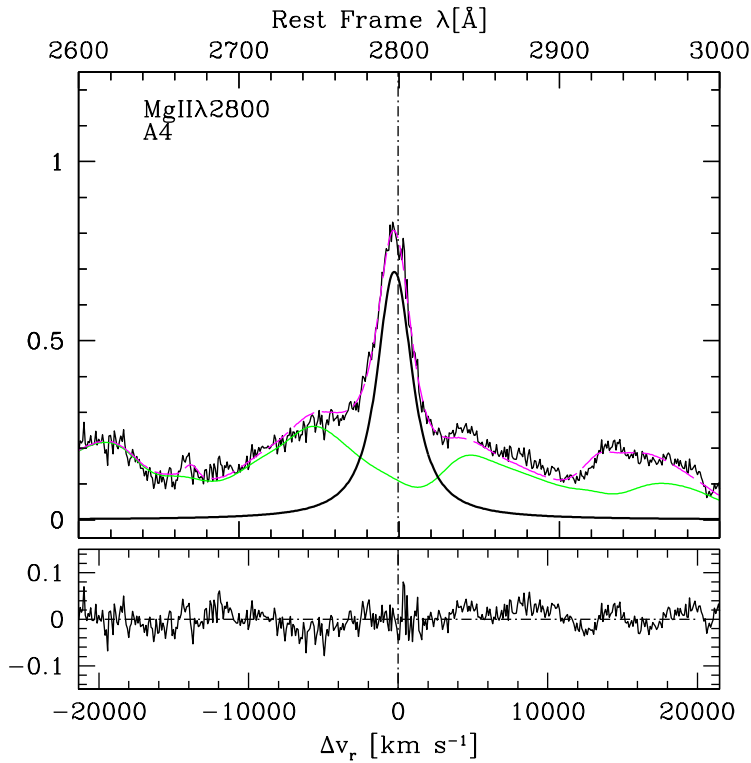}
\caption{The \mgii\ profile of the median spectrum for spectral type A4, from \cite{sulenticetal12}, after continuum subtraction. Abscissa scale is rest frame wavelength (top) and radial velocity (bottom) from rest frame wavelength of \mgii. Ordinate scale is continuum-normalized intensity. The dashed magenta line shows  a model fit  with all components included. The thick black line traces the  the \mgii\ doublet. The thin green line shows the assumed \feii\ emission. }
\label{fig:mgii}
\end{figure}

\section{A low-ionization wind/outflow}

\begin{figure}[ht!]
\centering
\includegraphics[width=5in]{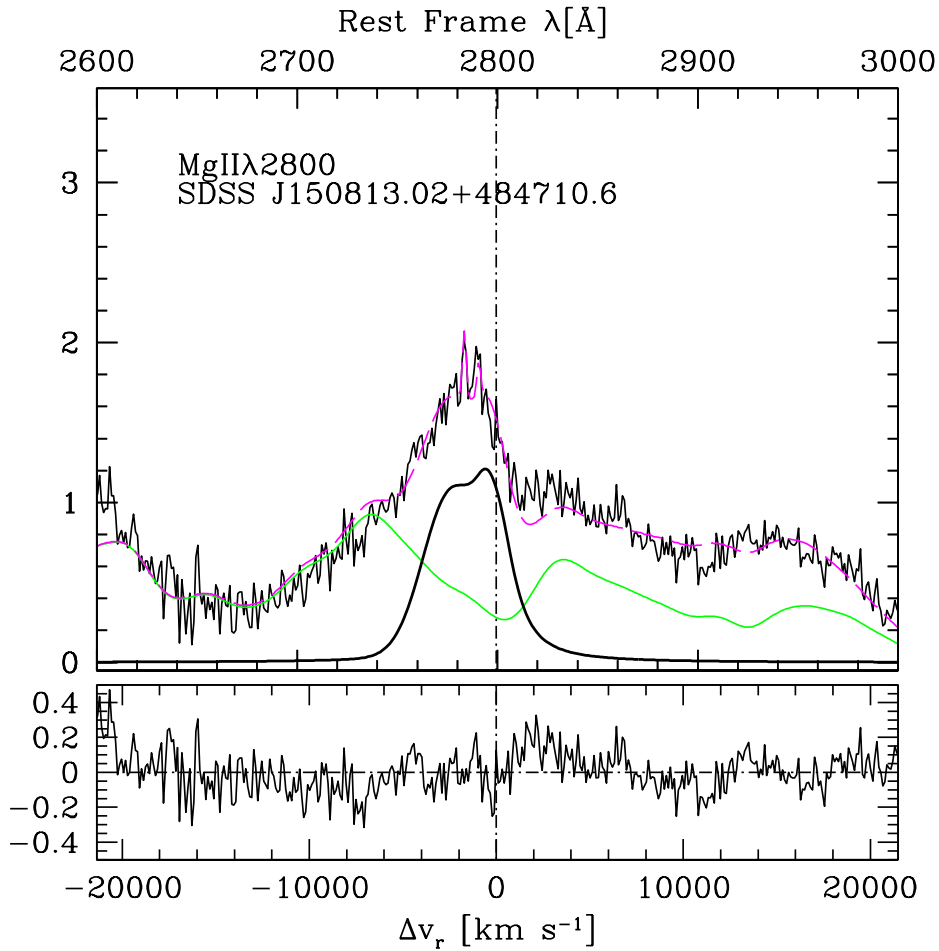}
\caption{The \mgii\ profile of the extreme Pop. A source SDSS J150813.02+484710.6. Meaning of symbol is as in the previous figure. Vertical scale is specific flux in units of 10$^{-15}$ \ergss\ \cmq\ \AA$^{-1}$.   Note the resemblance with the \civ\ profile of I Zw 1.}
\label{fig:sdss}
\end{figure}

Up to this point we have considered shifts of up to a few thousands \kms\ in the \civ\ line which is a high-ionization resonance line. \civ\ is assumed to be representative of HILs only for the simple reason that it is not strongly contaminated by other lines, is strong and conveniently placed.  Recent observations of smaller amplitude blueshifts in the LIL \mgii\ resonance doublet (Fig. \ref{fig:mgii}) came as a surprise \cite{marzianietal12}. The shift is not easy to measure because it depends on the intrinsic doublet ratio (that is only approximately known) and to a lesser extent on the estimate of the underlying \feii\ emission. A small shift by $\approx 200$ \kms\  would be overlooked in most previous studies (it was found in \cite{tytlerfan92}, although not at a significant level because of the large uncertainty). In addition to the small systematic shifts in bin A3 and A4 (see Fig. \ref{fig:mgii}), \cite{marzianietal12} found a source where the \mgii\  profile is shifted by $\approx 2000$ \kms. The \mgii\ profile of  SDSS J150813.02+484710.6 resembles a typical \civ\ profile for A3 and A4 sources. (Fig. \ref{fig:sdss}).

Apart from this extreme SDSS source we note that \mgii\ is different from both \civ\ and \hb.  In  \mgii\ we see a displacement of the {\em core} of the line.  The shift normalized by line width is large, with the implication that the line emitting gas is predominantly in outflow (see Fig. \ref{fig:mgii}).  What is the relation between the \mgii\ and \civ\ blueshifts?   In a photoionization framework \mgii\ emission is associated with low-ionization, and high column density gas that allows for the existence of a partially-ionized  zone \cite{kk81,netzer80,grandiphillips79}.  The partially-ionized zone extends to much larger depth within a gas cloud or slab than the fully ionized zone  where  \civ\ is  emitted. Such high \nc\ ($\gtsim 10^{23}$ \cmq) gas has to exist if photoionization is at work: we are speaking of a very prominent line in all type 1 quasar spectra.  The second ingredient for \mgii\ emission involves the strong X-ray continuum that is typical of quasars and that is needed to create the partially-ionized zone \cite{kk81}. For bin A3 and A4, \lledd$\rightarrow$1, and the force multiplier ${\cal M}$\  is large enough to drive a wind to a velocity larger than the virial velocity if only the effect of the ionizing continuum is considered. The relatively low amplitude of the shift may therefore result from the combined effects of large distance and, especially, of large column density \cite{marzianietal12}. The large emitting radius may favor shielding from the strong UV continuum radiation close to the equatorial plane of the disk where large column density material could be preferentially located \cite[e.g.][]{chelouchenetzer03,gaskell09,gibsonetal09,wuetal12}.

 


We note that these results would imply that  \mgii\ is not suitable as a virial estimator for \mbh\ computations in extreme Pop. A bins (which are a minority in a large quasar sample, typically around 10\%\ \cite{sulenticetal12,zamfiretal10}). Also the smaller width of \mgii\ with respect to \hbbc\ \cite{wangetal11,sulenticetal12} implies that a calibration between \mbh -- luminosity -- FWHM different from the one of \hb\ should be applied.


\section{Conclusion}

The existence of non-relativistic outflows in quasars has left the realm of speculation. High-ionization lines are, at least in part, produced in outflowing gas. There is evidence that this is likely to be true within both the BLR and NLR. Inter-line shift analysis has proved a powerful tool to understand this aspect of quasar structure and kinematics. The \civ\ blueshift is associated with a wind component whose prominence increases with \lledd\ along the 4DE1 sequence. In bins A2, A3 and A4 the wind component is basically dominating HIL fluxes. At low Eddington ratio  the \civ\ profile is less affected. \hb\ and \civ\ line profiles are, if not correlated, occupying overlapping ranges in  parameter spaces \cite{marzianietal96,marzianietal10}. This basic scenario has been confirmed by recent work \cite{richardsetal11,wangetal11}. Recent developments point toward the possibility of resonant line driving as an explanation
for outflows at low \lledd.  

Major improvements will come from 2D reverberation mapping which is, in principle, capable of resolving different coexisting dynamical regimes within the BLR \cite{horneetal04,denneyetal09a}. The multiplexing ability of new generation spectrographs like X\-SHOO\-TER \cite{vernetetal11} allow one to simultaneously observe the strongest LILs and HILs with higher sensitivity and will  provide new data of quality suitable for the measurements of small inter-line shifts.  This may lead to a better understanding of the outflows especially in low \lledd\ radiators, as well as to  a self-consistent unified  scenario of quasars outflows.

\paragraph{Acknowledgements}
PM acknowledges  support by Junta de An\-da\-lu\-c\'\i a, through grant TIC-114 and the Excellence Project P08-TIC-3531, and by the Spanish Ministry for Science and Innovation through grants AYA2010-15169 for a ``sabbatical'' stay at IAA, whe\-re  this paper was written.

\clearpage

\bibliographystyle{apj} 

%

\end{document}